\documentclass{appolb}
\usepackage[dvips]{graphicx}
\usepackage{cite}

\begin{document}
\title{SEARCH FOR THE $\eta$-MESIC HELIUM AT COSY%
\thanks{Invited talk  at the Excited QCD 09,  8-14 February 2009,  Zakopane, Poland }%
}
\author{
Pawe{\l}~Moskal$^{\star,\$,}$\footnote{E-mail address: p.moskal@fz-juelich.de}~\\
 on behalf of the COSY-11 and WASA-at-COSY collaborations
\address{
$^{\star}$ Institute of Physics, Jagiellonian University, Cracow, Poland\\
$^{\$}$ Institut f\"ur Kernphysik and J{\"u}lich Center for Hadron Physics,\\
Forschungszentrum J\"ulich, J\"ulich, Germany\\
}
}
\maketitle
\begin{abstract}
We review status and perspectives of the search of the eta~-mesic helium at 
the cooler synchrotron COSY.
\end{abstract}
  
\section{Introduction}
In 1996 Yamazaki et al. discovered a bound state formed out of a
$\pi^-$  meson and a $^{207}$Pb nucleus~\cite{yamazaki}. 
The negatively charged pion was bound with the lead nucleus
by means of the Coulomb interaction
and the interplay between an attractive Coulomb and a repulsive strong potential 
lead to the very narrow state with the width of less than one~MeV.
Due to the nature of the binding we can consider the state as a pionic atom.
 
It is also conceivable that a neutral meson could be bound to a nucleus. 
In this case the binding is exclusively due to the strong interaction 
and hence such object can be called a {\it mesic nucleus}. 
Here the most promissing candidate
is the $\eta$-mesic nucleus since the $\eta$N interaction 
is attractive and since there exists a baryonic resonance (N$^*(1535)$) 
which couples predominantly
to the $\eta$N channel~\cite{pdg}. 
We may  picture~\cite{sokol0106005} the formation and decay of the
$\eta$-mesic nucleus
as the $\eta$ meson absorption by one of the nucleons and 
then its propagation in the 
nucleus
via  subsequent excitation of nucleons to the N$^*(1535)$ state 
until the resonance decays into the pion-nucleon pair 
which escapes from the nucleus.
Predicted values of the width of such states range 
from $\sim$7 to $\sim$40~MeV~\cite{osetPLB550,liu3,haideracta}.

The search of the $\eta$~-~mesic nucleus was conducted
in many inclusive experiments~\cite{bnl,lampf,lpi,jinr,gsi,gem,mami}
via reactions induced by pions~\cite{bnl,lampf}, 
protons~\cite{jinr,gem}, and photons~\cite{lpi,mami}.
Many  promissing indications 
of the existence of such an object
were reported~\cite{lpi,mami,gem},
but so far none was independently confirmed.  
Experimental investigations with higher statistical sensitivity
and the detection of the N$^*(1535)$ decay products are being continued at 
the COSY~\cite{wasaproposal,wasaacta}, 
JINR~\cite{jinr},
and  MAMI~\cite{kruscheacta} laboratories.
In this contribution we report on searches of the $\eta$~-~mesic helium 
in  exclusive measurements carried out at the cooler 
synchrotron COSY by means of the 
WASA-at-COSY~\cite{wasaproposal,wasaacta}  
and COSY-11~\cite{c11acta,c11meson08,c11nuclphys,c11aps}     
detector setups.

We consider the study of the $\eta$-mesic nuclei as
interesting on its own account, but  additionally  
it is useful for investigations of
(i) the $\eta$N interaction, 
(ii) the N$^*(1535)$ properties in nuclear matter~\cite{jido},  
(iii) the properties of the $\eta$ meson
in the nuclear medium~\cite{osetNP710}, 
and 
(iv) the flavour singlet component  
of the $\eta$ meson~\cite{bass}. 

\section{Indications for the existence of the $\eta$~-~mesic helium}
\begin{figure}[h] 
    \begin{center} 
        {\includegraphics[scale=0.300]{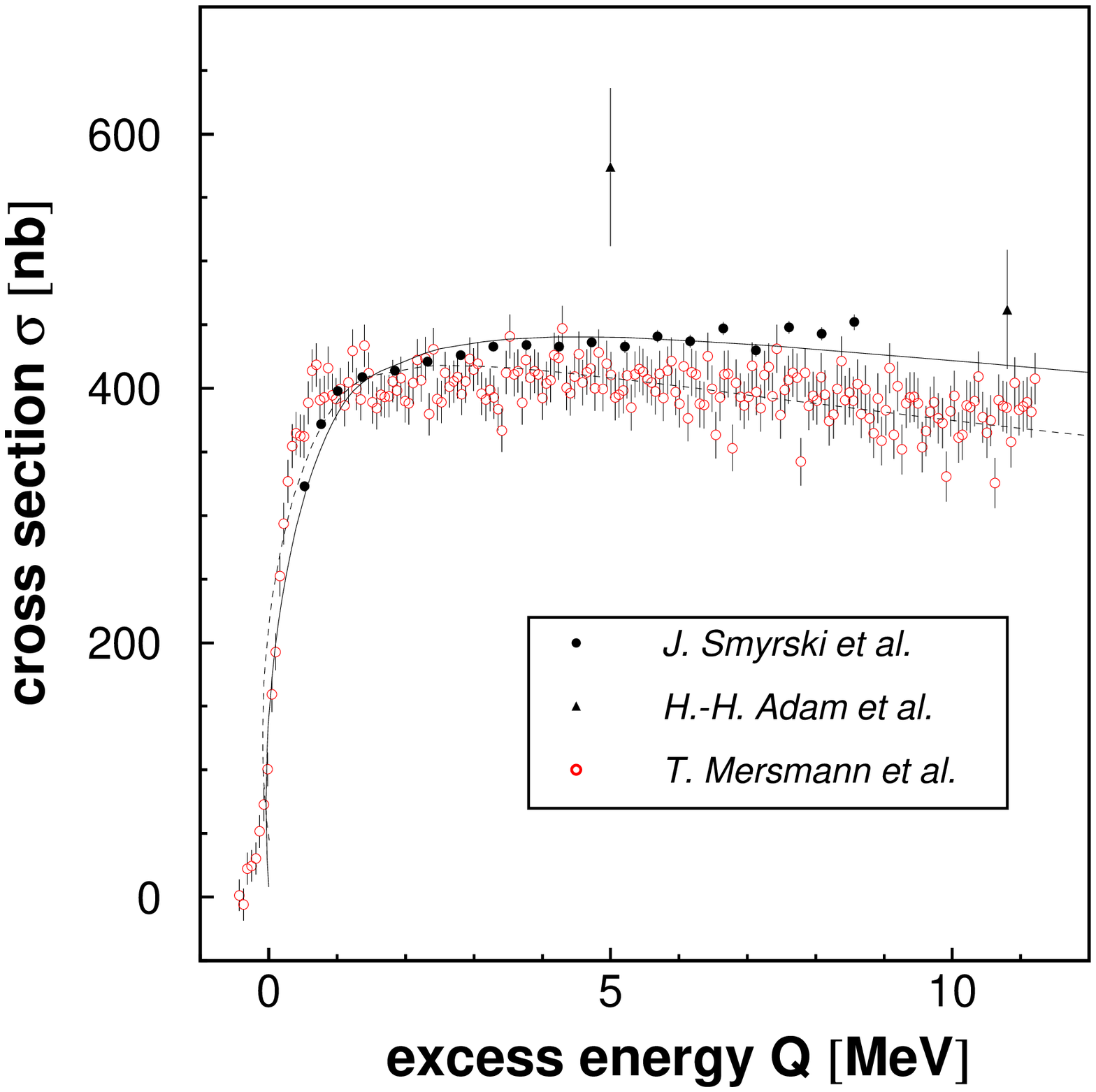}} \hfill
        {\includegraphics[scale=0.303]{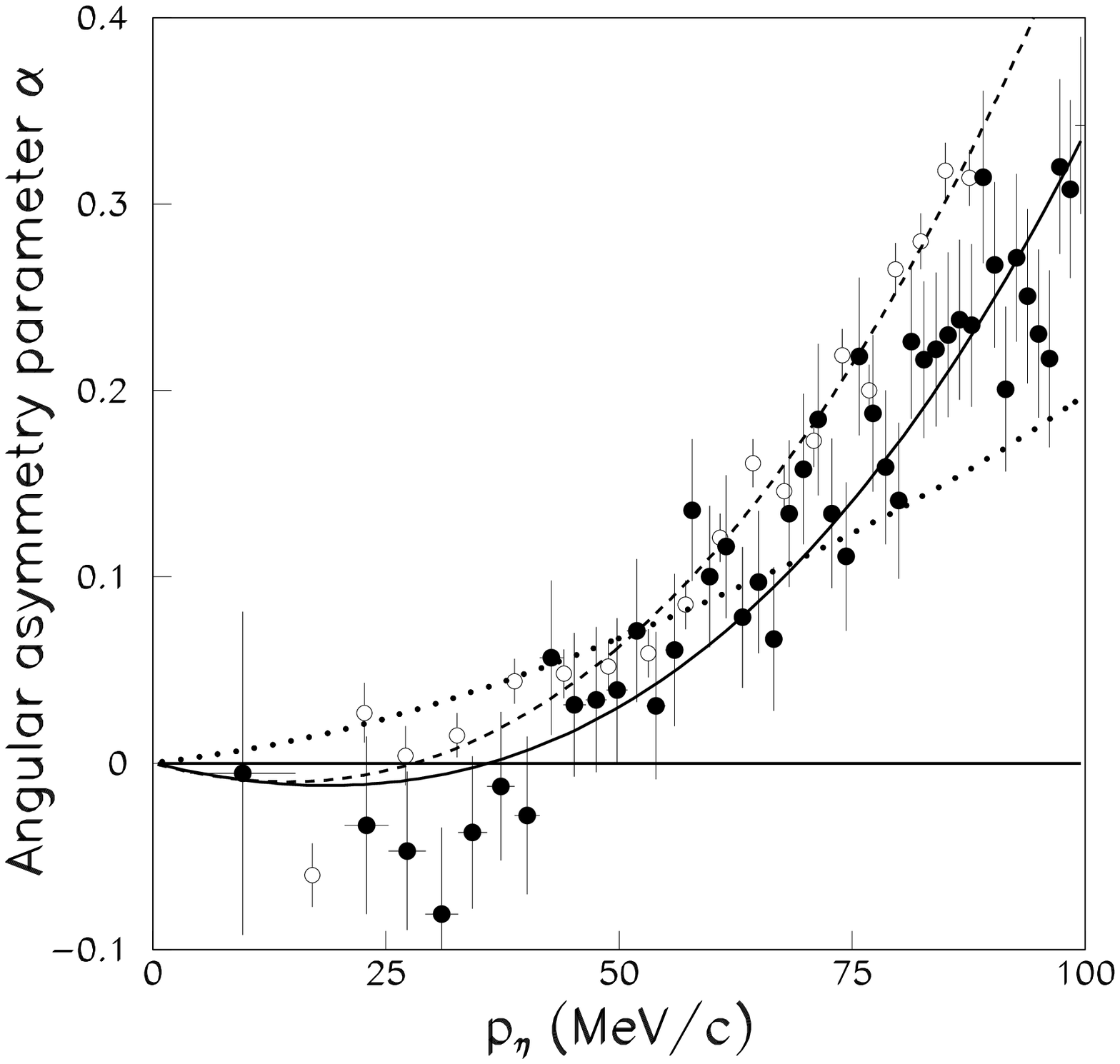}} 
        \caption{(left) Total cross section for the $dp\to ^{3}\!\!{\mathrm{He}}\eta$ reaction 
as determined by the
COSY-ANKE~\cite{timo} (open circles) and the COSY-11~\cite{jurek-he3} 
(full dots) and~\cite{adam} (triangles). The solid and dashed lines represent the scattering length fit 
to the COSY-11 and COSY-ANKE data, respectively.
(right) Angular asymmetry parameter $\alpha$. Closed and open circles represent results of
COSY-ANKE~\cite{timo} and COSY-11~\cite{jurek-he3}, respectively.
The dashed and solid lines denote results~\cite{wilkin2} 
of the fit (allowing for the phase variation) to the COSY-11 and COSY-ANKE data, respectively.
The dotted line denotes result of the fit  without the  phase variation.
The figure is adopted from Ref.~\cite{wilkin2}.} 
\label{fig1} 
    \end{center} 
\end{figure}

In 1985 Bhalerao and Liu~\cite{liu1} performed a coupled-channel 
analysis of the $\pi {\mathrm{N}}\to \pi {\mathrm{N}}$, 
$\pi {\mathrm{N}}\to \pi \pi {\mathrm{N}}$ and $\pi {\mathrm{N}}\to \eta {\mathrm{N}}$ reactions 
and discovered that the interaction between the nucleon and the $\eta$ meson
 is attractive. 
Based on this finding Haider and Liu  postulated the existence of the 
$\eta$--mesic nuclei~\cite{liu2}, in which the electrically neutral $\eta$ meson 
might be bound with the nucleons by the strong interaction. 
The formation of such a bound state can only take place if
the real part of the $\eta$-nucleus scattering length is negative 
(attraction), and the 
magnitude of the real part  is greater than the 
magnitude of the imaginary part~\cite{liu3,sibirtsev}. 
In the 1980's  
the $\eta$--mesic nuclei were considered
to exists for A~$\ge$~12~only~\cite{liu2} 
due to the  
relatively small value of the $\eta$N scattering length
($a_{\eta {\mathrm{N}}}~=~(0.28\, +\, i0.19)$~fm~\cite{liu1}).
However, recent theoretical investigations of hadronic- and photo-production 
of the $\eta$ meson result in values of $a_{\eta {\mathrm{N}}}$ which depending on the
analysis method range from
$a_{\eta {\mathrm{N}}}~=~(0.25\, +\, i0.16)$~fm up 
to $a_{\eta {\mathrm{N}}}~=~(1.05\, +\, i0.27)$~fm~\cite{wycech}, 
and which do not exclude the formation 
of a bound $\eta$-nucleus states  for such light nuclei as helium~\cite{wilkin1,wycech1} 
or even for deuteron~\cite{green}.
According to the calculations including multiple scattering theory~\cite{wycech1}
or Skyrme model~\cite{scoccola} an especially good candidate for binding 
is the $^4{\mathrm{He}} -\eta$ system.
Recent calculations by Haider~\cite{haideracta} or Tryasuchev and Isaev~\cite{tryasuchev} 
also indicate the binding in the $^4{\mathrm{He}} -\eta$ system,
while they rather exclude the existence of the $^3{\mathrm{He}} - \eta$ state. 
On the other hand there are promissing experimental signals 
which may be interpreted as indications of the the $^3{\mathrm{He}} -\eta$ bound state.
For example the shape of the excitation function 
for the $d\,p\to ^3\!\!{\mathrm{He}}\,\eta$ reaction~\cite{wilkin1},
determined by the SPES-4~\cite{berger}, SPES-2~\cite{mayer}, 
COSY-11~\cite{jurek-he3}, and COSY-ANKE~\cite{timo} collaborations~(Fig.~\ref{fig1}(left)). 
It has been indicated by Wilkin~\cite{wilkin2,wilkinaip} that a steep rise of 
the total cross section in the very close-to-threshold region followed 
by a plateau may be due to the existence of a pole 
of the $\eta ^3{\mathrm{He}}\to \eta ^3{\mathrm{He}}$ 
scattering amplitude in the complex excess energy 
plane $Q$ with $Im(Q)<0$~\cite{wilkin2}. This reference shows that the 
occurence of the pole changes the phase and the magnitude of the s-wave production amplitude. 
And indeed the momentum dependence of the asymmetry in the angular distributions 
of $\cos\theta_{\eta}$ expressed in terms of a
parameter $\alpha$ ~\cite{jurek-he3}, 
can only be satisfactorily described (solid and dashed lines in Fig.~\ref{fig1} (right)) 
if a very strong phase variation associated with the pole is included 
in the fits~\cite{wilkin2,wilkinaip}. 
Otherwise there is a significant discrepancy between the experimental data and 
the theoretical description (dotted line in Fig.~\ref{fig1} (right)). 

\section{Search of the $\eta-^3{\mathrm{He}}$ state with the COSY-11 detector setup}
Details of the experimental technique and the comprehensive description 
of results concerning the search of the $\eta$-mesic $^3{\mathrm{He}}$ nucleus  
conducted by the COSY-11 group 
were described elsewhere~\cite{c11acta,c11meson08,c11nuclphys,c11aps}.
Therefore, here we only briefly summarize the main results.
The measurements were carried out
using 
the deuteron beam of COSY which was circulating through 
the stream of the internal hydrogen target
of the cluster-jet type~\cite{brauksiepe}.
Data were taken during a slow  acceleration of the beam
from 3.095~GeV/c to 3.180~GeV/c, crossing the kinematical threshold
for the $\eta$ production in the $dp \rightarrow {^3{\mathrm{{He}}}}\,\eta$ reaction
at 3.141~GeV/c.
The determined excitation function for the  $dp \rightarrow {^3\mbox{He}}\,\pi^0$
process does not show any structure
which could originate from a decay of $\eta-{^3\mbox{He}}$ bound
state. The estimated upper limit for the cross section of the 
$dp \to (\eta^3\mbox{He})_{bound} \to {^3\mbox{He}}\,\pi^0$ 
reaction chain is equal to 70~nb~\cite{c11nuclphys}.
Similarly, the analysis of the $pd\to ppp\pi^-$ reaction resulted only in an
upper limit of 270~nb for the total  cross section 
of the $dp \to (\eta^3\mbox{He})_{bound} \to ppp\pi^-$ reaction~\cite{c11acta}.

\section{Search of the $\eta-^4{\mathrm{He}}$ state with WASA-at-COSY}
The installation of the WASA detector at COSY opened a unique possibility to search for
the $^4{\mbox{He}}-\eta$ bound state with  high statistics and high acceptance.
We have conducted a search 
via an exclusive measurement
of the excitation function for the $dd \rightarrow {^3\mbox{He}}\, p\, \pi^-$ reaction
varying  continuously
the beam momentum
around the 
threshold for the $d\,d \to ^4{\mathrm{He}}\,\eta$ reaction.
Ramping of the beam momentum and 
taking advantage of the large
acceptance of the WASA detector\footnote{For the coincidence registration of all ejectiles
from the $dd \to (\eta\,{^4\mbox{He}})_{bound} \to {^3\mbox{He}}\, p\, \pi^-$ 
reaction the acceptance of the WASA-at-COSY detector equals to almost 70\%.} 
allows to minimize systematical uncertainities making
the WASA-at-COSY a unique facility~\cite{wasa1}  for such kind of exclusive experiments.
The  ${^4\mbox{He}}-\eta$ 
bound state should manifest itself as a resonant like structure
below the threshold for the $dd \to {^4\mbox{He}}\,\eta$ reaction.
If a peak below the ${^4\mbox{He}}\,\eta$ threshold
is found, then
the profile of the 
excitation curve  will allow 
to determine the binding energy and the width 
of the $^4{\mathrm{He}}-\eta$ bound state.
If, however, only an enhancement around the threshold is found, then  it will enable
to establish the relation between width and binding energy~\cite{osetPLB550}.
Finally, if no structure is seen
the upper limit for 
the cross section of the production 
of the $\eta$-helium nucleus 
will be set at few nanobarns.  
In addition, when searching for the signal of the $\eta$-mesic state
we may take advantage of the fact that 
the distribution of the relative angle between the $nucleon-pion$ pair
for the background 
(due to the prompt $dd\to {^3{\mbox{He}}}\, p\, \pi^-$ reaction) is much broader 
than the one expected from the decay of the bound state.
This is because the relative angle between the outgoing $nucleon-pion$ pair  originating 
from the decay of the N$^*(1535)$ resonance 
is equal to 180$^{\circ}$ in the N$^*$ reference frame and it is 
smeared only by about 30$^{\circ}$ in the reaction center-of-mass frame 
due to the Fermi motion of the nucleons inside the $\mbox{He}$ nucleus.
This allows for an additional control of the background 
by comparing excitation functions corresponding to the "signal-rich" and "signal-poor" regions. 

In the first experiment conducted in June 2008, 
we used a deuteron pellet target and the COSY deuteron 
beam with a ramped momentum corresponding to a variation of the excess energy 
for the $^4{\mathrm{He}}-\eta$ system from -51.4~MeV to 22~MeV.
At present the data are evaluated and some preliminary results were reported~\cite{wasaacta}.
The experiment will be continued in
2010~\cite{wasaproposal}.
Two weeks of COSY beamtime  were already recommended by the COSY Program Advisory Committee. 

\section{Search of the bound state with quasi-free beams}
The systematic uncertainties in establishing the shape of the excitation functions
discussed in the previous sections are significantly reduced 
by using the momentum ramping technique since
the energy range of interest for the search of the $\eta$-mesic nucleus is scanned in each COSY cycle.
A scan of the energy can also be achieved in the case of the fixed beam energy
available for the external beam experiments. This may be realised by means 
of the quasi-free reactions as already successfully used at COSY for the study
of the meson production in quasi free proton-neutron collisions~\cite{prc2009,tof}.
For instance, a search of the  $\eta$-mesic Tritium may be realized by studying the exitation 
function of the $n\, d \to (\eta\, T)_{bound} \to d\, p\, \pi^-$   reaction
using a deutron beam and tagging the nd reactions
by the measurement of the spectator protons ($p_{sp}$) from the 
$d\, d\to p_{sp}\, n\, d\to p_{sp}\, (\eta\, T)_{bound} \to p_{sp}\, d\, p\, \pi^-$ reaction.
The Fermi motion of nucleons inside the deuteron beam will allow to scan a large 
range (in the order of 100~MeV) of excess energies in the nd reaction.

\section{Acknowledgements}
The work was 
supported by the
European Community-Research Infrastructure Activity
under the FP6 program (Hadron Physics,RII3-CT-2004-506078), by
the German Research Foundation (DFG), 
by the FFE grants from the Research Center J{\"u}lich,
and by
the Polish Ministry of Science and Higher Education 
(3240/H03/2006/31  and 1202/DFG/2007/03).

\end{document}